# Mercurian Impact Ejecta: Meteorites and Mantle

Brett GLADMAN* and Jaime COFFEY

University of British Columbia, Department of Physics and Astronomy
6244 Agricultural Road, Vancouver, British Columbia  V6T 1Z1, Canada
• Corresponding author. E-mail: gladman@astro.ubc.ca

## Abstract

 We have examined the fate of impact ejecta liberated from the surface of Mercury due to impacts by comets or asteroids, in order to study (1) meteorite transfer to Earth, and (2) re-accumulation of an expelled mantle in giant-impact scenarios seeking to explain Mercury's large core.  In the context of meteorite transfer during the last 30 Myr, we note that Mercury's impact ejecta leave the planet's surface much faster (on average) than other planet's in the Solar System because it is the only planet where impact speeds routinely range from 5—20 times the planet's escape speed; this causes impact ejecta to leave its surface moving many times faster than needed to escape its gravitational pull.  Thus, a large fraction of mercurian ejecta may reach heliocentric orbit with speeds sufficiently high for Earth-crossing orbits to exist immediately after impact, resulting in larger fractions of the ejecta reaching Earth as meteorites.  We calculate the delivery rate to Earth on a time scale of 30 Myr (typical of stony meteorites from the asteroid belt) and show that several percent of the high-speed ejecta reach Earth (a factor of 2—3 less than typical launches from Mars); this is one to two orders of magnitude more efficient than previous estimates.  Similar quantities of material reach Venus.
  These calculations also yield measurements of the re-accretion time scale of material ejected from Mercury in a putative giant impact (assuming gravity is dominant).  For mercurian ejecta escaping the gravitational reach of the planet with excess speeds equal to Mercury's escape speed, about one third of ejecta re-accretes in as little as 2 Myr.  Thus collisional stripping of a silicate proto-mercurian mantle can only work effectively if the liberated mantle material remains in small enough particles that radiation forces can drag them into the Sun on time scale of a few million years, or Mercury would simply re-accrete the material.

## Introduction

There are now several dozen lunar and martian meteorites.  While this represents  only a tiny fraction of the worldwide meteorite inventory, these meteorites are especially interesting due to their coming from a highly-evolved parent body.  Study of the cosmic-ray exposure (CRE) histories of these meteorites show that they have spent 0.6-20 Myr in space for the case of Martian meteorites, with a more compressed 0-10 Myr range for the lunar meteorites (see the recent review by Eugster et al 2006). The time scales turn out to be the natural dynamical transfer time scale, and numerical simulations of the orbital evolution of these meteorites (Gladman et al 1996, Gladman 1997) are in excellent agreement with CRE data, supporting the hypothesis that such meteorites are fragments that are simply delivered to Earth after being launched during hypervelocity impacts on the Moon and Mars.

While the lack of meteorites from Venus is understandable due to the screening effect of the venusian atmosphere, Mercury is an obvious candidate source for meteorites. The launch speed needed to escape the gravity well (4.2 km/s) is intermediate between that of the Moon (2.4 km/s) and Mars (5.0 km/s), so escape from the planet's surface is obviously feasible. Once escaped from the surface of Mercury, conservation of energy requires that the meteoroid's speed $v_\infty$ relative to Mercury after reaching `infinity' (in practice anything more than 100 planetary radii) is

$$v_\infty = \sqrt{v_{ej}^2 - v_{esc}^2} \quad , \quad (1)$$

where the ejection speed $v_{ej}$ from Mercury's surface must of course be larger than the planetary escape speed $v_{esc}$. Because of this relation, ejection speeds must exceed roughly twice the escape speed before the velocity relative to Mercury (which is then vectorially added to Mercury's heliocentric velocity to produce the escaping particle's heliocentric velocity) becomes comparable to the escape speed. This is rare in the lunar and martian cases, and results in the escaped meteoroids having heliocentric orbits very similar to their parent planet.

After escape into heliocentric orbit, the mercurian meteoroids orbit the Sun and will either impact a planet, impact the Sun, or perhaps be destroyed collisionally by other heliocentric projectiles. For mercurian meteoroids with $v_\infty < 8$ km/s (meaning ejection speeds < 9 km/s), the liberated particles are on orbits which only cross that of Mercury because Mercury's large average orbital speed of 48 km/s (which the ejected particles inherit on average, modified vectorially by the value of $v_\infty$) is difficult to change. Thus, the resulting orbits are very `Mercury-like', causing the most likely fate to be re-accretion onto the parent (Gladman et al 1996). In this regime Wetherill (1984) and later Love and Keil (1995) concluded on the basis of Monte Carlo orbital-evolution calculations that only 1 in 10,000 of the escaping mercurian meteorites reach Earth (about 100 times fewer than the martian meteorites). Gladman et al (1996) and Gladman (2003) performed full N-body calculations of the problem which took into account orbital resonant effects and found 0.1-0.5% of the mercurian meteoroids would reach Earth for $v_\infty = 2-6$ km/s, and concluded that for these yields it might not be surprising that we have no mercurian meteorites in the world-wide collection. (Note that we use the term `mercurian meteorites' instead of `hermean meteorites', due to the literature's general preference for the former). In this paper we argue that these low values for $v_\infty$ may be unrealistically underestimating the typical speeds and we re-calculate the transfer efficiencies for a variety of launch speeds.

# Launch Speeds at Mercury

Mercury is deep in the Sun's gravity well, and thus not only is Mercury's orbital speed high, but so are those of other objects who cross its orbit. In particular, asteroids and comets will cross the orbit of Mercury (and impact it) with speeds of many tens of km/s. Marchi et al (2005) examined the impact speed distribution of planet-crossing asteroids (which are in rough steady state due to their continual escape from the main asteroid belt), and concluded that most asteroidal impacts occur with impact speeds from 20 – 70 km/s. While the fractional contribution of large (>100 m diameter) comets as mercurian impactors is very unclear due to the potential for destruction as their perihelia are lowered down to ~0.4 AU, their larger semimajor axes (and often inclinations) cause the cometary impact speeds to be even higher (with speeds of 80 km/s and up).

These high impact speeds place Mercury in a unique position for planets in the Solar System, as it is the only planet which impactors strike at 5 – 20 times the planetary escape speed. In contrast, Earth is struck mostly at 15-30 km/s (1.5 – 3 times its escape speed) and 90% of the asteroids (the dominant impactor population) that strike Mars do so at speeds less than 4 times its escape speed (W. Bottke, private communication 2008). Since the maximum impact-ejecta launch speeds are a fraction of the impactor speed (up to about half for the theory of Melosh (1984), as an example) values of $v_\infty$ = 10 – 30 km/s are quite feasible for mercurian ejecta. This regime has not been quantified in earlier calculations.

# Method

In order to calculate the dynamical fate of mercurian ejecta, we have performed numerical integrations of the orbital evolution of large numbers of test particles escaping Mercury. For each simulation, test particles were randomly positioned moving radially away from Mercury with a single value of $v_\infty$; since in reality a single impact will not eject particles in all directions these initial conditions give results averaged over all impact locations on the planet. Using the same method as in Gladman et al (1996) we then tracked the ejecta's heliocentric orbits taking into account the gravitational influence of the planets (Mercury through Neptune), and stopped the integration of a particular particle if it struck a planet, had its perihelion reach the surface of the Sun, or reached 60 AU from the Sun (with the latter fate never happening in these simulations). The integrations were carried out for 30 million years of simulated time. The dynamics were purely gravitational although we discuss non-gravitational effects below.

**Numerical Simulations**

We studied the cases of $v_\infty$ = 4, 9, 14, 20, and 25 km/s by integrating 7000, 7000, 16 000, 18 000, and 18 000 test particles (respectively) for up to 30 Myr after their launch (the 14 km/s case was only integrated for 20 Myr). The variable number of particles was simply due to the available computer resources on the LeVerrier beowulf cluster at UBC at the time of the integration's start. In each case the number of launched particles is sufficiently large that the planetary impacts reported below are directly observed in the integrator, as opposed to estimating impact probability using the test-particle histories in conjunction with an Opik collision probability estimate (see Dones et al 1999). The simulations required roughly 30 CPU-years of computational effort.

The basic orbital evolution of the particles is characterized by repeated gravitational encounters with the terrestrial planets, each of which modifies the meteoroid's heliocentric orbit. Each planetary encounter could result (with low probability) in an impact instead of a flyby, which the integration algorithm then logs. The number of meteoroids in space will decline with time, and the region of orbital parameter space occupied by the meteoroids will expand as their orbits diffuse via gravitational slingshot effects with the planets into new regions of parameter space. For mercurian ejecta this allows meteoroids to reach larger semimajor axes as they climb out of the Sun's gravitational well due to fortuitous encounters with Mercury, Venus, and Earth (the opposite effect can also happen). The analysis presented here will focus on the distribution of impact times (after launch from Mercury) to Earth, Venus, and Mercury, expressed as a cumulative fraction (called the transfer efficiency) of the launched meteoroids that reach that target.

## Calculation Results

**Meteoroids reaching Earth**

We begin our analysis by examining the fraction of mercurian ejecta that reaches the Earth (and thus may become meteorites). In Fig. 1, we first note that the results of the 4 km/s case are in agreement with the previous calculations of Gladman (2003), namely that about 0.1% of mercurian meteoroids would reach Earth in the first 10 million years after launch. Because the CRE ages of stoney meteorites (including martian meteorites) are often factors of several longer than this, we have integrated three times longer (to 30 Myr) than the previous study and find an efficiency of 0.7% in 30 Myr for the 4 km/s case, again comparable to the 0.5% efficiency in 23 Myr estimated by Gladman et al (1996). This reproducibility (despite using different initial conditions, different machines, and a slightly different version of the integration code) provides confidence in the calculations. The lack of Earth impacts near the start of the simulation is due to the fact that the 4 km/s case has no ejecta initially crossing the Earth's orbit, or even that of Venus; gravitational scatterings by Mercury to reach Venus-crossing orbits then allow Venus scatterings to transfer some meteoroids to Earth-crossing orbits on time scales of 5 Myr.

Increasing $v_\infty$ to 9 km/s causes some of the ejecta to initially be on Venus-crossing orbits. This planet's greater mass allows it to rapidly assist meteoroids to Earth-crossing orbits. The rate of Earth impacts is roughly constant, with a cumulative efficiency of about 0.6%/10 Myr.

Further increasing $v_\infty$ to 14 km/s enters the regime in which some ejected meteoroids (those that were `traveling forward' relative to Mercury once liberated) to be in the Earth-crossing regime. Particles with aphelion distances slightly greater than 1 AU have high Earth-encounter probabilities per unit time (see Fig. 5 of Morbidelli and Gladman 1998) and these objects have high probabilities of being accreted by Earth on 10 Myr intervals.

Once $v_\infty$ = 20 – 25 km/s is reached, the Earth impact efficiency declines with further increases in launch speed. This is because many of the particles that are launched outwards at now deeply crossing Earth's orbit, which in fact lowers their per-orbit collision probability.

**Meteoroids reaching Venus**

Although the mercurian material accreted by Venus will likely be completely ablated in the venusian atmosphere, we present the results in Fig. 2 for completeness. Here the two slowest cases have similar yields, with Venus absorbing 20% of material escaping Mercury in 30 Myr; in these cases many of the Venus-crossing meteoroid orbits only cross in a shallow sense while at aphelion, which is the highest collision probability per orbit state. By 14 km/s the largest semimajor axis orbit cross Venus more deeply (lowering their collision probability per orbit, and they begin to enter the regime favorable for Earth impacts. This continues for larger speeds; for $v_\infty$ = 20 and 25 km/s the transfer efficiency to Venus drops to 13% and 8% in 30 Myr, respectively. These large yields (factors of several above those to Earth) simply reflect that Venus has a non-negligible cross-section and intersects a large fraction of the meteoroid swarm at any time.

**Meteoroids re-accreted by Mercury**

Since all of the ejecta is initially Mercury crossing, it is not surprising that it is Mercury itself that accretes the largest fraction of its ejecta. The efficiency and the time scale over which this occurs are important to understand in the context of Mercury re-accreting its own mantle if the planet's original outer layers were blown of in a giant collision.

For $v_\infty = v_{esc} = 4$ km/s the effect of gravitational focusing enhances the re-accretion rate until a combination of gravitational scattering to higher relative speeds and the preferential elimination of high collision-probability meteoroids via accretion slows the re-accretion rate (see Gladman et al 1995 for a more complete discussion in the case of lunar ejecta). This results in fully half of the ejecta being re-accreted by Mercury in 30 Myr. Since a further 20% has been accreted by Venus (figure 2), only about one third of the ejecta still survives in space after this time. It is likely that at least half of this remaining material would be swept up by Mercury in the following tens of Myr.

The gravitational focusing effect is much reduced in the 9 km/s case. For larger ejection speeds Mercury's gravitational focusing becomes unimportant, with all the higher ejection speeds having roughly 25% of the launched material returning to the planet. Taking into account the material absorbed by Venus and Earth, about two thirds of the mercurian ejecta ejected with $v_\infty = 20 - 25$ km/s is still in space 30 Myr after launch.

## Discussion

The transfer efficiency to Earth shown by these simulations (2 – 5% in 30 Myr) is an order of magnitude larger than previous estimates, and is about half of the martian delivery efficiency (cumulative martian efficiencies are shown in Gladman 1997). However, we must first address the issue of whether radiation effects could invalidate the purely gravitational approach we have taken here.

**Poynting-Robertson drag and Yarkovsky drift**

Particles in the inner regions of the Solar System are subjected to intense solar radiation; the role of the resulting non-gravitational forces may need to be considered when evaluating the orbital evolution of the mercurian ejecta. Although solar radiation is responsible for several forces, the motion of centimeter to decimeter-sized particles is predominantly affected by Poynting-Robertson drag (Robertson, 1937). In the particle's reference frame, the Sun's intercepted radiation is absorbed and re-emitted isotropically. In the solar frame of reference, however, more momentum is lost in the forward direction. This results in a velocity-dependent force which acts in the direction opposite to the particle's velocity, decreasing the meteoroid's heliocentric semimajor axis *a*. The time scale for the orbital evolution to bring a circular orbit with initial semimajor axis *a* down to the Sun can be shown to be (see Wyatt and Whipple 1950):

$$T_{sprial} = \frac{7\,Myr}{Q_{pr}} \left(\frac{\rho}{1\,g/cc}\right)\left(\frac{r}{1\,cm}\right)\left(\frac{a}{1\,AU}\right)^2 \qquad (2)$$

where $\rho$, *r*, and *a* are the meteoroid's density, radius, and orbital semimajor axis respectively, and $Q_{pr} \simeq 1$ is the Poynting-Robertson drag coefficient for this size range (Burns et al. 1979). When eccentricity becomes important numerical integration of the spiraling timescale from the PR-evolution equations has been presented in tabular form by Wyatt and Whipple (1950). Taking *e* = 0.5 and *a*=0.7 AU for a typical outward-flung meteoroid with density of 3 g/cc and radius of 10 cm (similar to martian meteoroids), the PR collapse time scale is ~70 Myr. Thus, except for the meteoroids with *a*<0.5 AU (which are less likely to reach Earth-crossing orbits), PR drag should be relatively ineffective at dragging decimeter-scale mercurian meteoroids down to the Sun; in addition, our simulations show that meteoroids which are on Earth and/or Venus-crossing orbits tend to be scattered by encounters with those planets to larger semimajor axes and perihelia, further weakening the PR effect.

A second non-gravitational perturbation is the Yarkovsky effect, in which the higher momentum carried away from the hotter afternoon region of the body causes a systematic acceleration (see Bottke et al 2006 for a recent review). For decimeter-radius meteroids the so-called 'diurnal' effect dominates and produces a maximum drift rate ~0.03 AU/Myr (Vokrouhlicky 1999 and private communication 2008). This rate is not negligible (order unity change in semimajor axis over 15 Myr), but the drift could be outwards (for prograde rotators) or inwards (for retrograde rotators), and thus Yarkovsky could help meteoroid delivery to Earth by pushing about half of randomly-oriented impactors out towards Venus crossing orbits. The drift rate will not be monotonic if collisions with small particles in solar orbit can re-orient the spin axis. On average Yarkovsky may in fact be beneficial since reaching the Venus-crossing regime would be the rate-limiting step for low-speed ejecta, whereas the high-speed ejecta that is already on Earth-crossing orbits will again be prone to be scattered to larger semimajor axis and perihelion where Yarkovsky effects will be less important.

**Mercurian meteorites on Earth**

This greater yield we calculate, enabled by the fast launches that likely occur from the mercurian surface, re-opens to the possibility that mercurian meteorites could be present in the worldwide collections. Love and Keil (1995) discussed the issue of what properties such meteorites would have and thus how we would recognize them as mercurian (as opposed to asteroidal). The surface reflectance spectra provided by ground-based studies (eg., Sprague et al 2007) and soon to be provided by the Messenger spacecraft (see Boynton et al 2007) will doubtless provide constraints on discussions of possible links of certain meteorite types to Mercury as a source body (examples: Palme 2002, Keuhner et al 2006, 2007, Markowski et al 2007).

The calculations presented here indicate that mercurian meteorites reaching Earth would most likely (1) have pre-atmospheric radii of order 1 decimeter and 4-pi CRE ages (for the transfer itself, as opposed to a potentially longer near-surface 2-pi CRE residence age) of 5-30 Myr. This is similar to martian meteorites with the interesting difference that the long CRE-age martian meteorites are reduced by collisional destruction in the main belt (Gladman 1997) whereas mercurian meteorites may be more limited by PR drag. Our calculations show that the atmospheric entry speeds of mercurian meteoroids will be faster (15-30 km/s) than is common for martian meteoroids, for which 10-15 km/s is typical; the mercurian entry speeds are more in line with `normal' chondritic meteorites and thus the lower ablation concluded from cosmic-ray records and explained by low-speed arrivals (Gladman 1997) should be absent for mercurian meteorites. Although unlikely that sufficient fireballs will be observed, incoming mercurian meteorites will exhibit a strong preference to fall in the morning rather than afternoon. Lastly, the common low masses of lunar meteorites will be rare in the mercurian case since cm-scale mercurian ejecta is rapidly depleted by PR drag.

**Mercury re-accretion efficiency and dispersing the proto-Mercurian mantle**

Re-accretion by Mercury is of relatively little interest for the meteorite problem, but is of great importance for quantifying the hypothesis that the high bulk density of Mercury is due to the collisional stripping of the mantle from a proto-mercurian planet by a giant collision (recently reviewed by Benz et al 2007). In this scenario the lower-density mantle of an already-differentiated proto-Mercury is ejected into space by the impact and then lost into the Sun via radiation effects.

The dominant uncertainty in this hypothesis is the departure speed and physical form of the escaping mantle material. Although not stated in giant collision studies, it is reasonable to assume that in cases where the mercurian core re-accretes but the mantle escapes, the escaping material will reach infinity with speeds comparable to the escape speed, and thus the $v_\infty$ = 4 km/s case of Fig. 3 is the most appropriate. In this case half of the material is re-accreted by Mercury in 30 Myr. In fact, in a similar recent study for 2 Myr (Benz et al 2007) those authors found that Mercury re-accreted just over 30% of its ejecta in 2 Myr; the good match with the 4 km/s case of Fig. 3 indicates that much of the ejected mantle does indeed depart with this speed.

However, as Benz et al (2007) noted, these re-accretion efficiencies and time scales assume that gravity is the only force in operation. Whether or not this is true depends on the size of the ejected particles and on the efficiency of PR drag in moving them down to the solar surface. Benz et al (2007) estimate that the cooling mantle ejecta will condense into almost exclusively 1 – 5 cm radius spheres, which we take to have density ~2.5 g/cc. Eq. (2) implies that low-e particles with $a$~0.4 AU will have orbital collapse time scales ~ 3 Myr, comparable to their accretion time scale, with eccentric orbits collapsing several times more rapidly. Benz et al (1988, 2007) argued that the material could be disposed of into the Sun before it could be added back to Mercury (otherwise the iron fraction of the post-impact Mercury would not be raised by the temporary removal of the mantle).

Whether this convenient disposal of the stripped mantle will work depends critically on the conclusion regarding the size of the ejecta, for Eq. 2 shows that the spiraling-in time scale increases linearly with the radius. Take the stripped mantle (Benz et al. 2007) to have a mass of 1.25 Mercury's present mass and add the putative impactor of 0.35 Mercury's mass (1/6 of the target). Taking the typical post-impact orbital inclination to be of order $(v_\infty/v_{orb})/2$, this material will be spread out in a thin ring surrounding the Sun. Putting the ejected mass into 1-cm radius drops of density 2.5 g/cc and spreading them out over this ring, we calculate that their optical depth to the solar radiation is of order unity, and probably higher in the mid-plane. The ring is thus potentially capable of self-shielding the particles from the effects of PR drag, which will lengthen the destruction time scale. Benz et al (1998) point out that this ring of particles is likely collisional, and with encounter speeds of order $v_\infty$ = 4 km/s the collisions would disrupt these small particles rather than allowing accretion; the resulting smaller fragments will be even more prone to PR drag and further collisions. The collisional cascade will increase the surface area to mass ratio of the fragments and cause the optical depth perpendicular to the midplane to exceed unity, making the collision time scale drop to the orbital period. The ring will thus collisionally evolve much faster than PR drag can dispose of individual particles. Although there will be a phase in which the inner ring edge (which has passed optical depth unity) will shield the rest of the ring from solar radiation, this may not last once the ring becomes very thin due to the fact that the solar surface will subtend of order a degree and particles may still receive radiation from above and below the mid-plane. However, once the plane is this thin the inclinations and eccentricities will have damped away and relative speeds become tiny, opening the possibility that the ring could re-accumulate into planetesimals, most of which would then subsequently be swept up by Mercury. (It is fascinating to consider that this scenario would temporarily make life very dark for all the other planets in the Solar System as the ring would block essentially all of the outbound solar light until it is disposed of into the Sun or re-accumulates into larger planetesimals.) Clearly the evolution of this ring of material is a complex question.

It is clear that more details are needed regarding the exact physical state of the ejected mantle material. It is not yet clear that no re-accumulation of this material into clumps larger than 1 cm will occur during the low relative-speed departure from Mercury, nor is it clear exactly how this material will evolve in solar orbit in a dynamical and collisional sense. Although the disposal scenario is plausible, much work remains to be done about the dynamical/collisional evolution of this massive amount of material now liberated into heliocentric orbit.

## Conclusions

We have measured the delivery rates via numerical integration in the `gravity-only' regime for mercurian ejecta for 5 launch speeds and find that 40-70% of the ejecta is swept up by Mercury, Venus, and Earth within the first 30 Myr after launch. At low speeds Mercury re-accretion is dominant due to the gravitational focusing caused by that planet's gravity, with 1/3 of the ejecta returning in 3 Myr and ½ within 30 Myr. For ejected meteoroids leaving Mercury with $v_\infty$ speeds in excess of ~9 km/s, 2-5% of the ejecta reaches Earth within 30 Myr (a time scale over which typical stoney meteorites survive). This rate is sufficiently high that mercurian meteorites may be more plentiful than previously thought.

In the context of a mantle-stripping impact from Mercury, the calculations give the return rates for mercurian ejecta to the planet and thus the time scale over which radiation forces must remove the material from Mercury-crossing orbit. The temporary heliocentric ring of material created may be sufficiently self-interacting that it could re-accumulate after collisionally damping to a thin ring and be swept up again by Mercury.

**Acknowledgements**: The authors have benefited from discussions with E. Asphaug, W. Bottke, J.A. Burns, and D. Vokrouhlicky. This work was supported by funding from NSERC, CFI, and BCKDF.

**Figures**

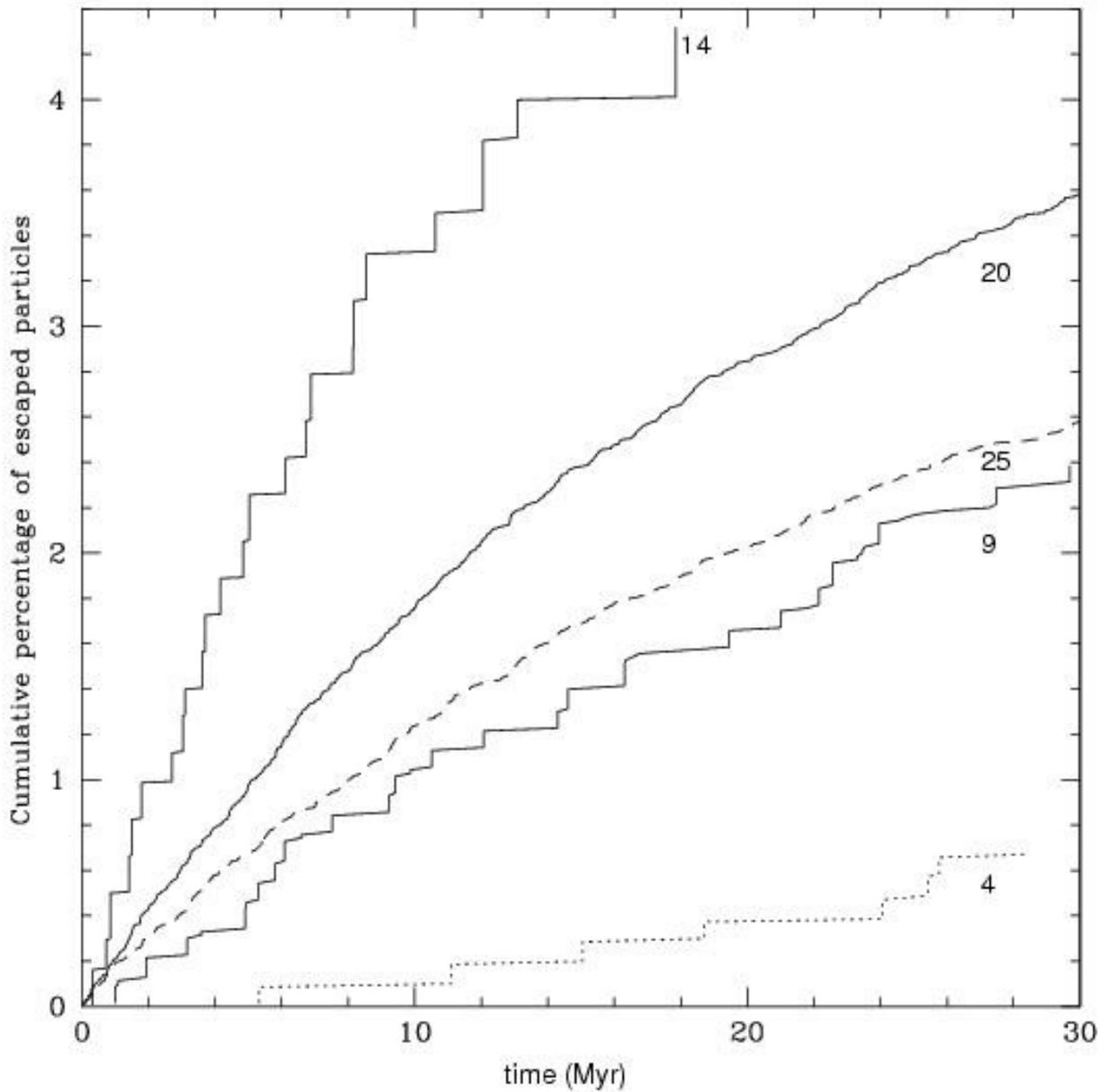

*Figure 1: Cumulative percentage of test particles ejected from Mercury impacting Earth, as a function of time after the ejection from Mercury. The value of $v_\infty$ used (in km/s) for the escaping test particles is indicated. There are no immediate Earth impacts for the 4 and 9 km/s cases since no particles are initially placed on Earth-crossing orbits. The smaller percentage of Earth impacts from the 20 and 25 km/s cases relative to the 14 km/s case occurs because the ejected particles can initially be placed on deeply-Earth crossing orbits (see text for discussion).*

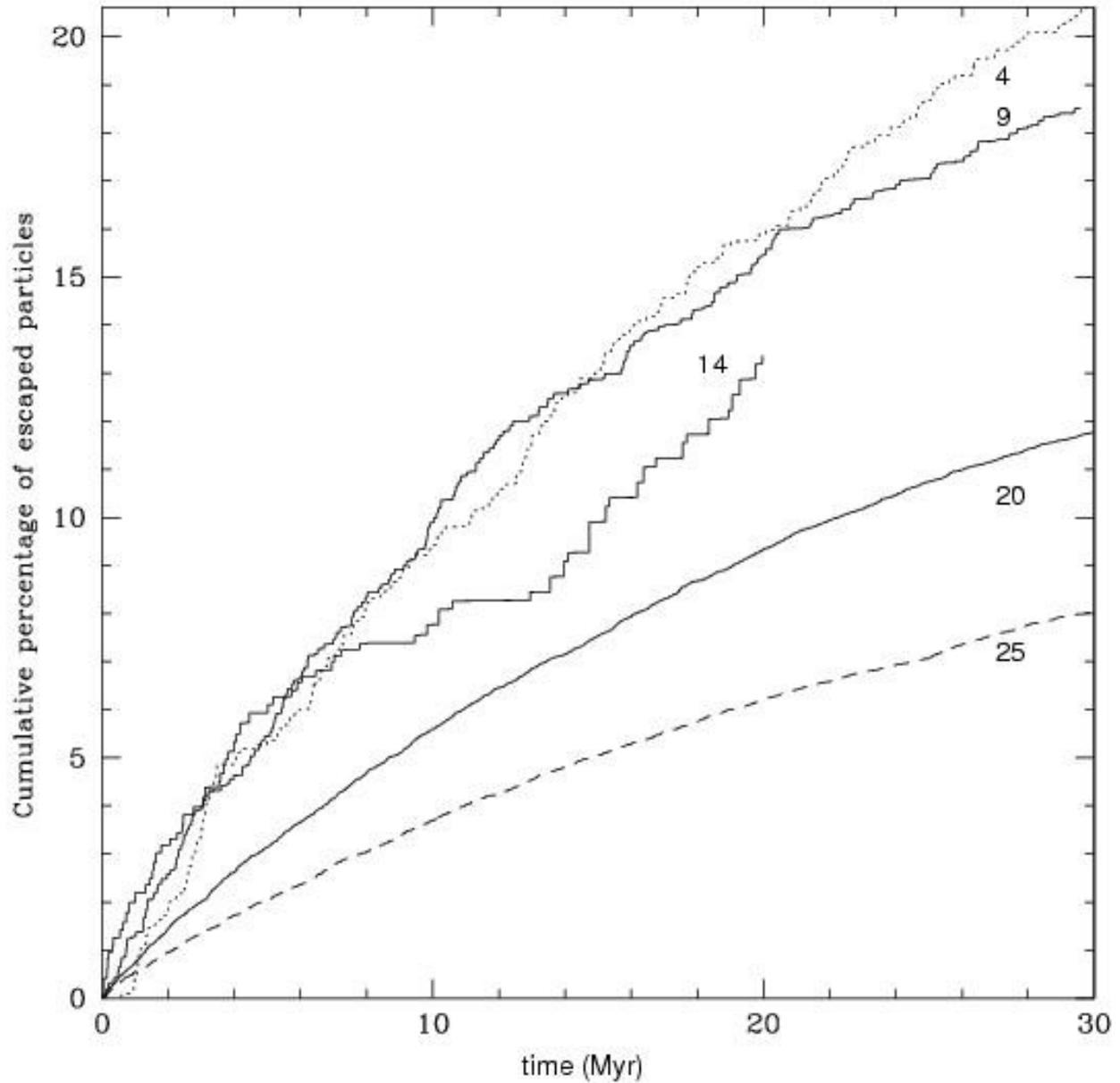

*Figure 2: Cumulative percentage of test particles ejected from Mercury impacting Venus, as a function of time after the ejection from Mercury. Note that the initial lack of impactors in the 4 and 9 km/s cases seen in figure 1 is essentially absent here, as it requires negligible time for some meteoroids to reach Venus-crossing orbits.*

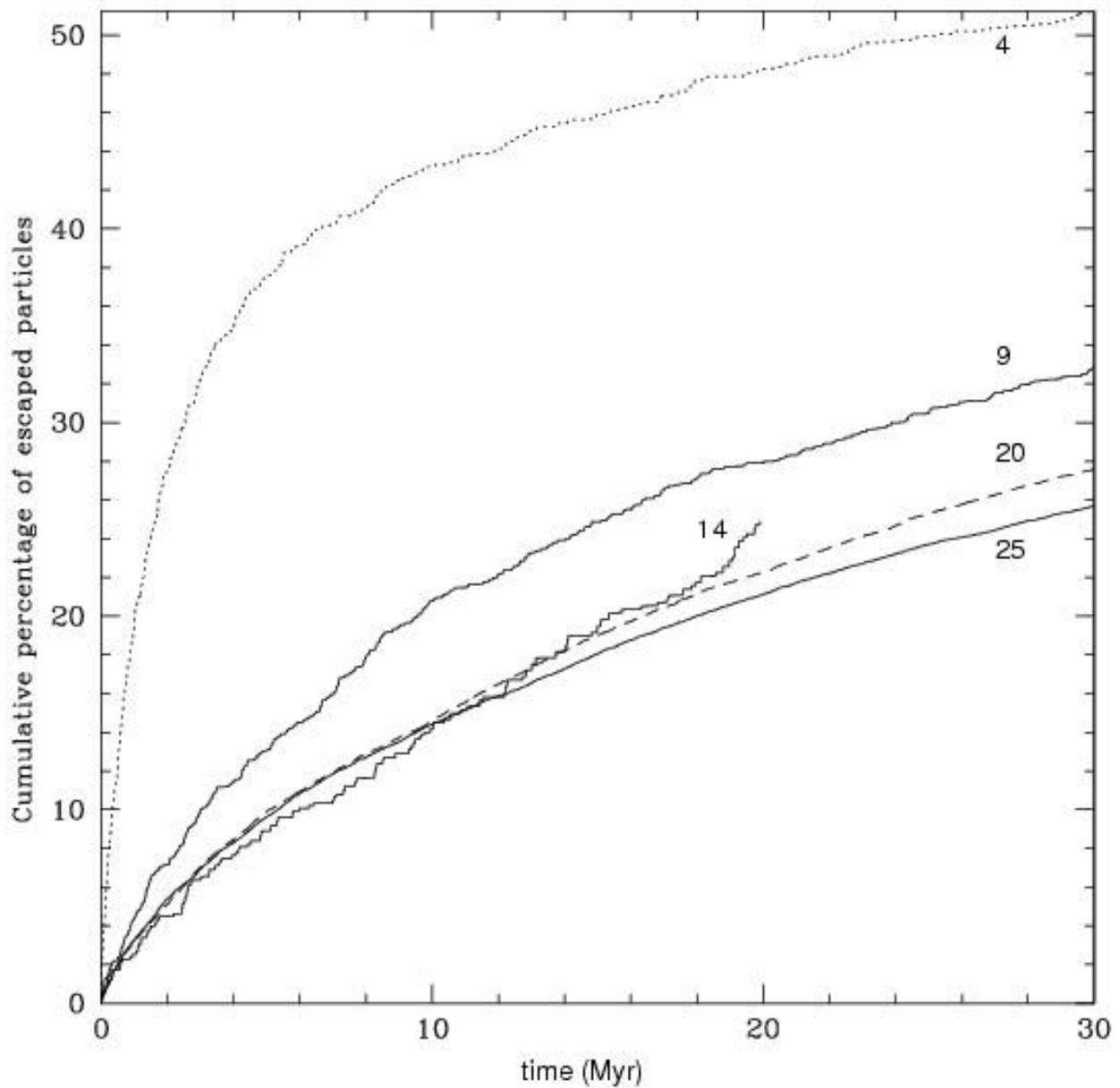

*Figure 3: Cumulative percentage of the test particles that re-impact Mercury in the purely gravitational problem, as a function of time after ejection from Mercury. Since the ejecta is initially on Mercury-crossing orbits, large amounts of matter is accumulated by Mercury in all cases. As discussed in the text, gravitational focusing causes half of the 4 km/s ejecta to be re-accreted by Mercury in 30 Myr.*